\documentclass[aps,twocolumn,showpacs]{revtex4}

\usepackage{graphicx}
\usepackage{dcolumn}
\usepackage{bm}
\usepackage{amssymb}

\begin{document}
\title{Fractal boundary basins in spherically symmetric $\phi^4$ theory}

\author{Ethan Honda}
\affiliation{
Falls Church, VA 22043
}
\email{ehonda@alum.mit.edu}

\date{\today}

\begin{abstract}
Results are presented from numerical simulations of
the flat-space nonlinear Klein-Gordon equation with an asymmetric double-well potential
in spherical symmetry.
Exit criteria are defined for the simulations that are used to help understand the 
boundaries of the basins of attraction for Gaussian ``bubble"  initial data.
The first exit criteria, based on the immediate collapse or expansion of bubble radius, 
is used to observe the departure of the scalar field from a static 
intermediate attractor solution.
The boundary separating these two behaviors in parameter space
is smooth and demonstrates a time-scaling 
law with an exponent that depends on the asymmetry of the potential.
%
The second exit criteria differentiates between the creation of an expanding 
true-vacuum bubble and dispersion of the field leaving the false vacuum;
the boundary separating these basins of attraction is shown to demonstrate fractal behavior.
The basins are defined by the number of bounces that the field undergoes before 
inducing a phase transition.
A third, hybrid exit criteria is used to determine the location of the boundary to arbitrary precision
and to characterize the threshold behavior.
The possible effects this behavior might have on cosmological phase transitions 
are briefly discussed.

\end{abstract}


\maketitle

\section{\label{sec:Intro} Introduction}

While expanding and collapsing scalar field bubbles have each been
given a great deal of attention in the physics community 
(largely in the contexts of phase transitions and oscillons, respectively), 
there is surprisingly little focus on the solutions that live on the threshold of those two end states.
This paper focuses on the critical phenomena that exist near this threshold 
and the fractal boundaries  that define it. 

The model used in this paper is  the nonlinear 
Klein-Gordon (nlKG) equation with an asymmetric double-well potential (ADWP).
The ADWP used here is the typical quartic potential with two stable minima: one 
 global minimum at the lowest possible potential energy (the ``true" vacuum) 
 and one local minimum at a higher energy (the ``false" vacuum).
 In the study of phase transitions, one typically considers initial data that consist of 
 perturbations on top of the false vacuum.  
 If the perturbations are large enough, the 
 tendency for the field to transition to the energetically more favorable true vacuum can 
 overcome the effective ``surface tension" that would cause the bubble to collapse.  
 When that is the case,
 the bubble wall will expand outward causing a transition from the false to the true vacuum;
 the resulting solution is referred to as a supercritical bubble.  
 If the perturbations are small, however, a phase transition will not occur; 
 the scalar field will collapse and eventually disperse,
 and the solution is referred to as a subcritical bubble.

When the nlKG equation was first studied in the context of phase transitions, 
research efforts focused largely on supercritical bubbles, and in particular, on domain walls.
Subcritical bubbles tended to be neglected because, by definition, they do not  
induce a phase transition;
it was not understood until much later that subcritical bubbles could still have 
cosmological implications.  
%
For a wide range of initial data leading to subcritical bubbles,
there exist localized oscillating solutions, called oscillons, that have lifetimes that 
are large compared to naive expectations. 
Oscillons were originally discovered by Bogolubsky and Makhankov
\cite{Bogo_Paper} and have been thoroughly investigated by numerous authors
using many types of nonlinear models
\cite{Honda_Paper,GF_PF_ZH_AL_SmallQBs,CGM_Oscillons,GH_Oscillons_HeatBath,
MG_LongLivedAsymmetricBubbles, GF_PF_ZH_MM_OscillonRadiation,
MG_RCH_Resonant_2p1, Farhi_SU2_Oscillons, Hindmarsh_Salmi_2006, Fodor_QuasiBreathers2006,
NGraham_Electroweak};
the effect oscillons may have on phase transitions  has been of great interest
\cite{MG_BR_JT_BubblingAway, MG_JT_U1, MG_JT_HiggsComplexity}. 
The work by Gleiser and Sicilia \cite{MG_DS_AnalyticOscillons} is an excellent 
summary and analytic characterization of the  salient features of oscillons.

While most of the simulations presented in this paper result in the formation of
either an oscillon or a domain wall, this paper actually focuses on neither;
this paper focuses on the nonlinear
dynamics that exist on the threshold of creating one of these two end states.  
%
%
This paper begins with a presentation of general formalism and definitions 
in Section \ref{sec:Formalism}.  
In Section \ref{sec:IntermediateAttractor}, methods are presented that 
can be used to solve for the intermediate attractor solutions to the
nlKG-ADWP system.  It is shown that in the neighborhood of the attractor,
Type I critical phenomena similar to those observed in gravitational collapse
are present.
Section \ref{sec:FractalBoundary} demonstrates that the basin boundaries
have fractal structure and 
can be parameterized by the number of times they ``bounce" before 
inducing a phase transition, not unlike the kink/antikink soliton (domain 
wall) collisions of  \cite{Matzner_Paper}.  
Furthermore, methods are presented that allow for the arbitrarily precise determination
of the location of the fractal edges. 
Being able to explore these edges in parameter space reveals that both
a static intermediate attractor
solution and Type I critical phenomena appear to exist on every basin boundary.
Finally, Appendix  
\ref{app:MinkowskiDimension} 
presents a rough approximation 
for the Minkowski-Bouligand fractal dimension of a data set with the basic properties 
observed in the nlKG-ADWP system.

\section{General Formalism and Definitions \label{sec:Formalism}}

The nlKG  action is
\begin{equation}
S[\phi ] = \int d^4x \sqrt{|g|}\left( - \frac{1}{2} g^{\mu\nu}
\partial_\mu\phi \, \partial_\nu\phi 
-V(\phi)
\right),
\label{eqn:action}
\end{equation}
where $\phi\equiv\phi(r,t)$ and $g_{\mu\nu}$ is the flat spacetime metric on a 
four-dimensional manifold.  The potential $V$ is given by
$V(\phi) = \frac{1}{4}\phi^2\left(\phi^2-4(1+\delta)\phi +4\right)$,
where  $\delta$ is a measure of the asymmetry of the potential.
When $\delta = 0$, $V(\phi)$ is a symmetric double-well potential with vacuum
states at $\phi=0$ and $\phi=2$.  With $\delta\neq 0$, the false and true vacuums
are $\phi_F = 0$ and 
$\phi_T = \frac{3}{2}\left(1+\delta\right)+\frac{1}{2}\sqrt{1+18\delta + \delta^2}$,
respectively.

Variation of the action with respect to $\phi$
gives the covariant equation of motion 
\begin{equation}
\frac{1}{\sqrt|g|}\partial_\mu\left(
\sqrt{|g|} g^{\mu\nu}\partial_\nu \phi \right) = 
\phi\left( \phi^2 - 3(1+\delta)\phi + 2\right),
\end{equation}
which after imposing spherical symmetry and 
using  spherical coordinates ($t$,$r$) gives
\begin{eqnarray}
\dot{\Pi} &=& \frac{1}{r^2}\partial_r\left( r^2 \Phi \right) 
-\phi\left(\phi^2 -3(1+\delta)\phi +2\right) \label{eqn:PiDot}\\
\dot{\Phi} &=& \partial_r\Pi \label{eqn:PhiDot}\\
\dot{\phi} &=& \Pi \label{eqn:phidot},
\end{eqnarray}
where $\Pi = \partial_t\phi$ and $\Phi_r = \partial_r\phi$.
The spacetime admits a timelike Killing vector that gives rise to the conserved energy
\begin{equation}
\label{eqn:energy}
E = 4\pi \int_0^{r_b} \!\!  dr \, r^2 \left[  \frac{1}{2}\left( \Pi^2 + \Phi^2 \right) + V(\phi) - V(\phi_T) \right],
\end{equation}
where $r_b$ is the radius of the outer boundary, and 
the constant $V(\phi_T)$ is subtracted  to give $E=0$ for $\phi(r,t) = \phi_T$ (i.e., everywhere at the 
true vacuum).

The initial data used to evolve 
(\ref{eqn:PiDot}), (\ref{eqn:PhiDot}), and (\ref{eqn:phidot})
are taken to be a simple Gaussian bubble:
\begin{equation}
\label{eqn:FreeSpaceInitialData}
\phi(r,0) = \phi_F + \left(\phi_T - \phi_F\right)
\exp\left( -r^2/\sigma_0^2 \right).
\end{equation}
This bubble type of initial data interpolates between the true and false vacuums and 
represents a perturbation away from the false vacuum that may (or may not) lead to a 
phase transition.

To explore the dynamics of the scalar field bubbles, 
it is useful to define an ancillary function
\begin{equation}
\label{eqn:xidef}
\xi(t_j) = \left\{
\begin{array}{ll}
\textrm{max} \left( r_i\left(\phi_{TF},t_j\right) \right) & \rm{when} \ \  \phi_{TF}\in\phi(r_i,t_j)  \\
0  & \rm{otherwise},
\end{array}
\right.
\end{equation}
where $\phi_{TF} \equiv (\phi_T + \phi_F)/2$, and $\phi_{TF}$ 
satisfies  $\phi_{TF} \in \phi(r_i,t_j)$
at $t_j$ if $\phi(r_i,t_j) \leq \phi_{TF} < \phi(r_{i+1},t_j)$ for some $i$.
More simply put, 
$\xi(t)$ is the maximum radius for which the field is halfway  between 
the true and false vacuums, and zero if at time $t_j$ the field does not anywhere
equal $\phi_{TF}$.
For example,
using the initial data profile (\ref{eqn:FreeSpaceInitialData}) and the definition 
(\ref{eqn:xidef}), one can define $\xi_0 \equiv \xi(t_0=0)$, 
which is simply the radius  (at half-maximum) of the Gaussian profile, $\xi_0 = \sigma_0 \sqrt{\ln 2}$.
For supercritical  bubbles, $\xi(t)$ is a reasonable approximation to the location of the bubble wall.

For collapsing bubbles, the number of oscillations, or bounces, of $\xi(t)$, 
denoted $N\left(\xi\left(t\right)\right)$, is defined to be the number of times
\begin{equation}
\begin{array}{rclc}
\xi(t_j) & = & 0 & \textrm{and  }\\
\xi(t_{j+1}) & > & 0, &  \\
\end{array}
\end{equation}
for any $j$. 
To retain common terminology with the works on kink/antikink 
soliton collisions, the term ``bounce" will be used to describe the oscillations
the field undergoes.
%
%

Finally, all solutions to  (\ref{eqn:PiDot}), (\ref{eqn:PhiDot}), and (\ref{eqn:phidot}) are obtained 
using a second-order iterative Crank-Nicholson 
finite difference scheme with higher-order Kreiss-Oliger dissipation \cite{Honda_Paper}.  
A simple outgoing boundary condition is used at the exterior of the grid, but the grid is chosen
to be large enough such that any reflected radiation could not travel back to origin during the
evolution.   Since this work is focused on the threshold of 
bubble expansion and not the long-term behavior of oscillons, 
this approach is simple and effective.

\section{The intermediate attractor\label{sec:IntermediateAttractor}}

\begin{figure}
\includegraphics[width=7.5cm,height=7.5cm]{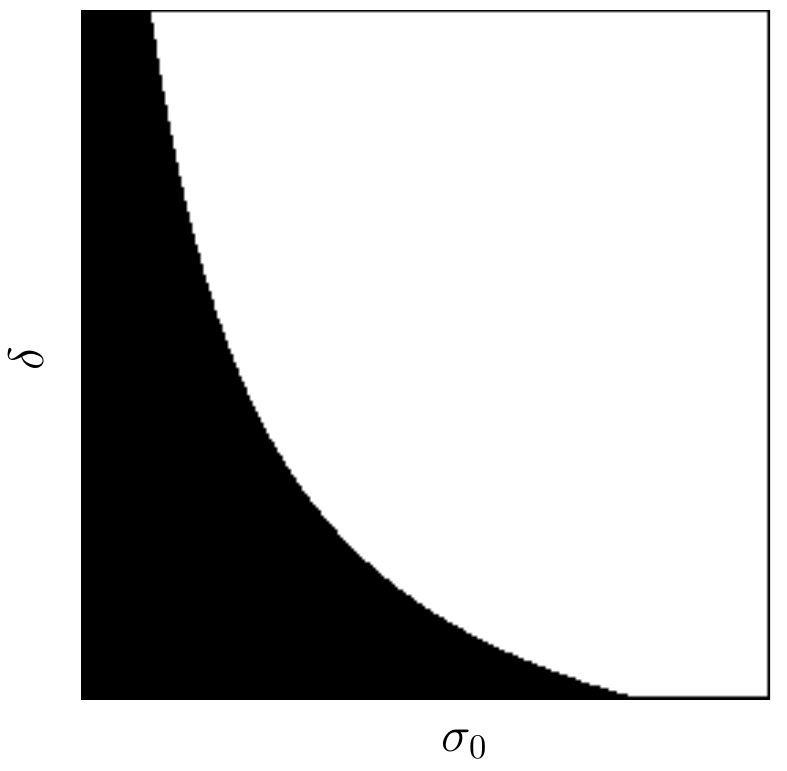}
\caption{
Bitmap of two-dimensional parameter space survey showing the collapse (black region)
or expansion (white region) of initial data with Gaussian width $\sigma_0$
and asymmetry parameter  $\delta$.
$\mathcal{E}_1$ is used to determine expansion or collapse
for 62\,500 (250$\times$250) pairs ($\sigma_0$,$\delta$), for 
$1.0 \leq \sigma_0 \leq 7.5$ and 
$0.05 \leq \delta \leq 0.25$.
The boundary between collapse and expansion is smooth.
}
\label{fig:simple_scaling}
\end{figure}

It has been shown in \cite{Honda_Diss} that  
solutions to  (\ref{eqn:PiDot}), (\ref{eqn:PhiDot}), and (\ref{eqn:phidot}) obey
a time-scaling law near the threshold of expansion or collapse.  
The exit criteria for the simulations in  \cite{Honda_Diss}  is based on the bubble
radius moving sufficiently far away from  the  initial bubble radius, $\xi_0$:
\begin{equation}
\mathcal{E}_1 =  \left\{
\begin{array}{lcrcl}
\textrm{ expansion} & \rm{when}   & \xi(t) & \geq & \xi^+   \\
\vspace{-2mm} \\
\textrm{ collapse }  & \rm{when}  & \xi(t) & \leq & \xi^-,
\end{array}
\right.
\label{eqn:Condition1}
\end{equation}
where $\xi^+$ and $\xi^-$ are constants chosen to be  above and below $\xi_0$, respectively.
Figure \ref{fig:simple_scaling} displays 
the  end state (expansion or collapse)  as a function of 
the width of the initial bubble $\sigma_0$ and the asymmetry parameter $\delta$.
The boundary between the two end states is observed to be smooth,
and the $\delta$-dependent threshold value of $\sigma_0$ is defined to be $\sigma^*_0(\delta)$, 
or just $\sigma^*_0$.

\subsection{Time-scaling law \label{sec:TimeScaling}}

To explore the properties of the threshold, one must find $\sigma^*_0$ precisely.  This 
is done by fixing the asymmetry parameter $\delta$ and varying
$\sigma_0$.   The search begins with a $\sigma^+_0$ that 
is known to expand and a $\sigma^-_0$ that is known to collapse.  
Evolutions are run with test values that successively bisect $\sigma^+_0$ and $\sigma^-_0$
until machine precision is reached  (64-bit double-precision variables used).
For a fixed asymmetry parameter $\delta$, one can observe the lifetime of the solution 
as a function of $\sigma_0$.  
The solution lifetimes are observed to obey a  time-scaling law
\begin{equation}
T = -\gamma \ln | \sigma_0 - \sigma^*_0 |,
\label{eqn:TimeScaling}
\end{equation}
with an exponent $\gamma$ that depends on $\delta$.  
\begin{figure}
\includegraphics[width=8cm,height=8cm]{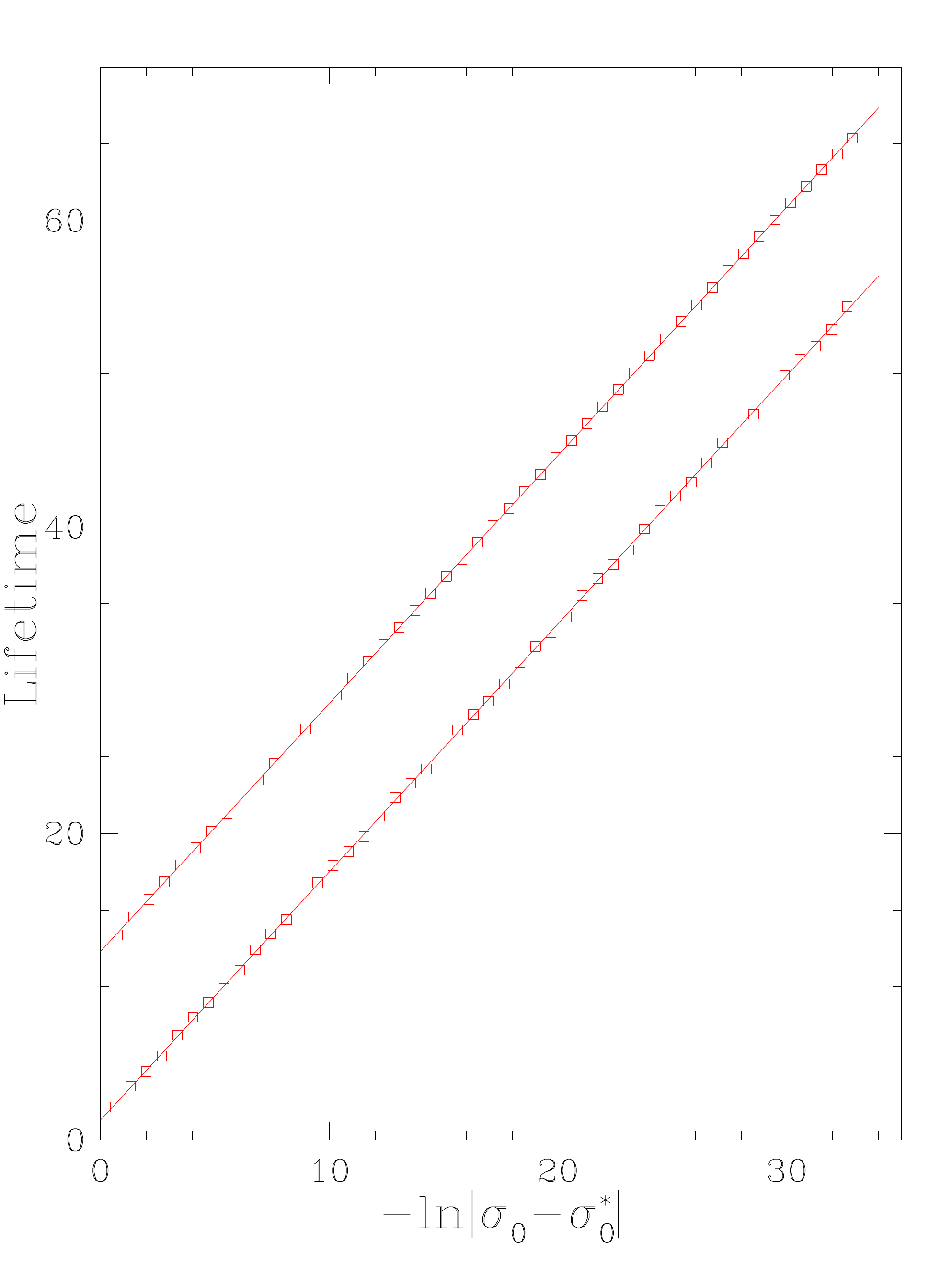}
\caption{ 
Plot of solution lifetime
as a function of $-\ln| \sigma_0 - \sigma^*_0|$
for  $\delta = 0.141$ using exit criteria $\mathcal{E}_1$.
Time-scaling exponent $\gamma = 1.62$ is  observed for both expanding (top line)
and collapsing (bottom line) evolutions.
Similar scaling exponents 
$\gamma=\{3.94, 2.13, 1.54, 1.23, 1.04\}$ were measured for 
$\delta=\{ 0.05, 0.10, 0.15, 0.20, 0.25\}$, respectively.
 }
\label{fig:TimeScaling1}
\end{figure}
The time-scaling law can be observed
on both the expansion and collapse sides of $\sigma^*_0$ in Fig. \ref{fig:TimeScaling1}.

The time scaling laws like (\ref{eqn:TimeScaling}) are not uncommon  in the study
of nonlinear field theories.  Similar Type I critical phenomena are observed 
in the gravitational collapse of a wide range of matter fields
\cite{Matt_Paper,PRB_CMC_SMCVG_Phases, MWC_EWH_SLL_ApproxBH, MWC_TC_PB_EYM, 
RB_JM_EYM, PB_TC_Skyrmions, Gundlach_Review}.  
In these models, in the limit that $\sigma_0$ approaches $\sigma^*_0$,
the solutions each approach an unstable intermediate attractor solution
with  Lyapunov exponent $1/\gamma$.
While the exact nature of the intermediate attractor solutions varies depending on the physical model,
the most common attractor type is  static
\cite{RB_JM_EYM,MWC_TC_PB_EYM,PB_TC_Skyrmions,MWC_EWH_SLL_ApproxBH}. 

\subsection{ The nlKG static intermediate attractor}

\begin{figure}
\hbox{\hspace{-5mm}
\includegraphics[width=8.5cm,height=8.5cm]{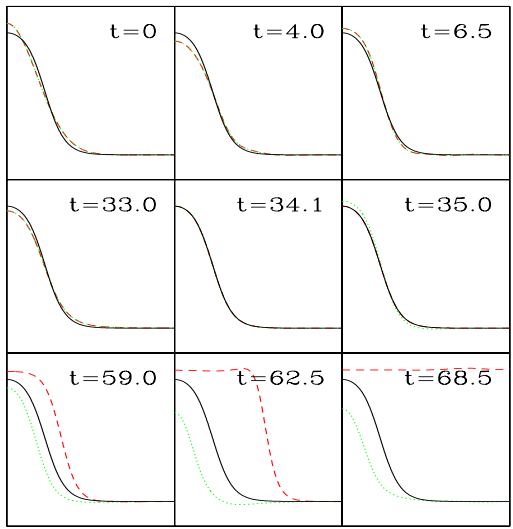}
}
\caption{(Color figure and movie online)
Time evolution of $\phi(r,t)$ for static (solid black), collapsing (dotted green),
and expanding (dashed red) solutions with $\delta = 0.141$, using $\mathcal{E}_1$.
Initial bubble radius $\sigma_0$ is
fine-tuned to $\sigma^*_0$ to within machine precision (64-bit floating point). 
The collapsing and expanding solutions oscillate about the static solution 
and disperse, leaving the spacetime in different vacuum states.
}
\label{fig:TimeEvolTiling1}
\end{figure}

Not surprisingly, observation of the threshold dynamic solutions shows that the field 
appears to consist of small oscillations on top of an underlying static solution.
Fig. \ref{fig:TimeEvolTiling1} shows snapshots of a near-critical 
evolution for values of $\sigma_0$ slightly above and below $\sigma^*_0$.  
Guided by this observation and other examples of Type I critical phenomena,
a static solution to (\ref{eqn:PiDot}), (\ref{eqn:PhiDot}), and (\ref{eqn:phidot})
is obtained by invoking a static ansatz (setting all time derivatives equal to zero).  
The static field equations are
\begin{eqnarray}
\frac{1}{r^2}\partial_r\left( r^2 \Phi \right) &=&
\phi\left(\phi^2 -3(1+\delta)\phi +2\right) \label{eqn:PhiPrime}\\
\partial_r{\phi} &=& \Phi, \label{eqn:phiPrime}
\end{eqnarray}
and are solved by ``shooting."
\begin{figure}
\includegraphics[width=8cm,height=8cm]{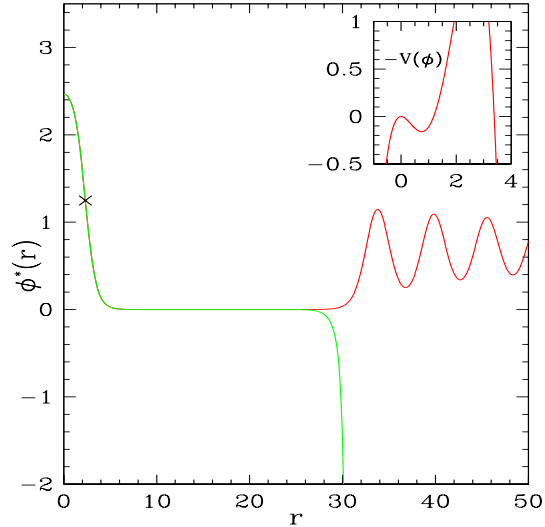}
\caption{(Color online)
Solution $\phi^*(r)$ obtained from the solution of (\ref{eqn:PhiPrime}) and
(\ref{eqn:phiPrime}) with $\delta=0.141$.  
This solution is the $\delta$-dependent static intermediate attractor
for (\ref{eqn:PiDot}), (\ref{eqn:PhiDot}), and (\ref{eqn:phidot}) at the 
collapse/expansion threshold fine-tuned to machine precision.
The ``x" denotes the value of $\xi \approx 2.3$.  
The inset shows $-V(\phi)$.
}
\label{fig:static_shoot1}
\end{figure}
\begin{figure}
\includegraphics[width=8cm,height=8cm]{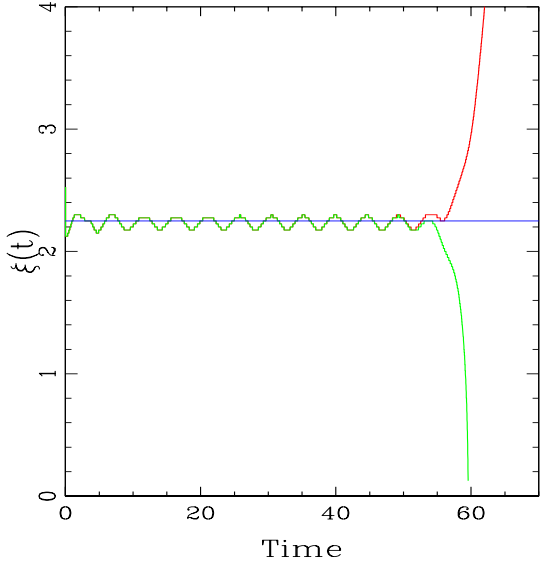} 
\caption{ (Color online)
Plot  of $\xi(t)$ obtained from evolving 
(\ref{eqn:PiDot}), (\ref{eqn:PhiDot}), and (\ref{eqn:phidot})
for two values of $\sigma_0$, one slightly above $\sigma^*_0$ and one slightly 
below for $\delta = 0.141$.  The solution with $\sigma_0 > \sigma^*_0$ leads to an expanding 
true-vacuum bubble.  The solution with $\sigma_0 < \sigma^*_0$ leads to
a bubble that collapses. 
The horizontal line represents the value of $\xi$ for the static intermediate attractor.
}
\label{fig:xi_tuned1}
\end{figure}

The solution of (\ref{eqn:PhiPrime}) and (\ref{eqn:phiPrime})
is very similar to the one-dimensional particle subject to a potential of
$-V(\phi)$, differing only by the geometric factor in the spatial Laplacian and 
the exchange of $r$ and $t$.
The critical solution is obtained by solving (\ref{eqn:PhiPrime}) and (\ref{eqn:phiPrime})
with $\Phi(r=0)=0$ and different initial values of $\phi(r=0)$.
The solutions have three types of behavior as  $r\rightarrow\infty$: 
$\phi\rightarrow \pm \infty$ or oscillation in the local minimum of $-V(\phi)$.
Bisecting on the threshold of the $\phi\rightarrow -\infty$ end state and the 
oscillation end state yields the desired critical solution that 
asymptotically approaches the false vacuum.  In the one-dimensional particle analogy, this is like 
finding the perfect initial position of the particle so it can ``roll" through the trough of the potential 
and come to a perfect stop at the unstable minimum at $\phi=0$ 
(Fig. \ref{fig:static_shoot1} inset).
Fig. \ref{fig:static_shoot1} shows the fine-tuned solution to (\ref{eqn:PhiPrime}) and 
(\ref{eqn:phiPrime}) for $\delta = 0.141$.  

In addition to showing the evolution of the dynamic solutions, Fig. \ref{fig:TimeEvolTiling1}
also shows the static intermediate attractor solution.  
It appears that the threshold solution is a superposition of the 
static solution and a small amplitude time-dependent localized ``shape mode," reminiscent 
of one-dimensional kink/antikink soliton collisions \cite{Campbell_Paper}. 
Fig. \ref{fig:xi_tuned1} shows the approximate bubble radius, $\xi(t)$, 
for the same two near-critical solutions shown in Fig. \ref{fig:TimeEvolTiling1};
it can be clearly seen that the bubble radius for the 
dynamic solutions oscillates about the bubble radius of the static intermediate attractor
before collapsing or expanding.

\section{True/False vacuum boundaries \label{sec:FractalBoundary}}

In the previous section it was shown that $\mathcal{E}_1$
divides the $\sigma_0$-$\delta$ parameter space into regions of  
immediate expansion or collapse. 
This section focuses on the more physical concern of what  vacuum state  
the spacetime approaches in the $t\rightarrow\infty$ limit.
%
In regions where the bubbles expand, $\xi>\xi^+$, the bubble wall
expands outward indefinitely and induces a phase transition to the true vacuum. 
In regions where the bubbles collapse, however,
$\mathcal{E}_1$ terminates the evolution immediately after $\xi(t)$ drops below $\xi^-$,
and as such,  no conclusion can be drawn about which vacuum state the field will 
occupy as $t\rightarrow\infty$. 
There are three long-term behaviors observed in the collapsing field configurations:
solutions that immediately disperse to leave the false vacuum, 
solutions that form  oscillons that eventualy disperse to leave the false vacuum, 
and solutions that collapse but bounce back and lead to an expanding true-vacuum bubble.

\begin{figure}
\includegraphics[width=8cm,height=8cm]{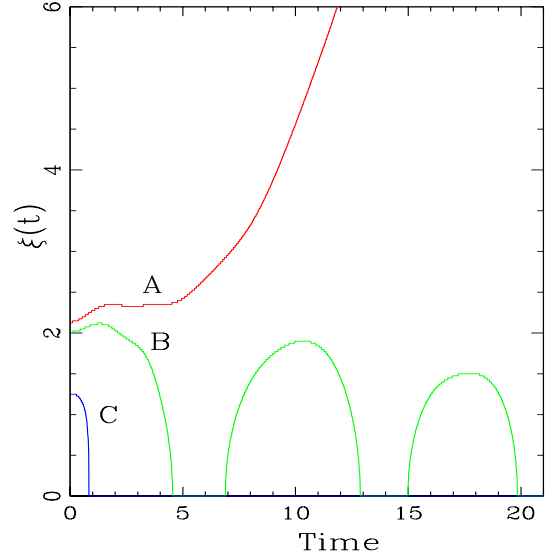} 
\caption{ (Color online)
Plot of $\xi(t)$ for three solutions subject to criteria $\mathcal{E}_2$.  Solution A is supercritical. 
Solution B is subcritical and exits due to  the $N(\xi(t)) > N_{\rm osc}$ condition
(only three oscillations are shown).
Solution C is subcritical and exits due to  the $t>t_{\rm max}$ condition.  Since solution C has so
little energy, it does not oscillate even once, i.e., $N(\xi(t)) = 0$ for all $t$.
}
\label{fig:exit2}
\end{figure}
\begin{figure}
\hbox{
\hspace{-1mm}
\includegraphics[width=6.9cm,height=6.9cm]{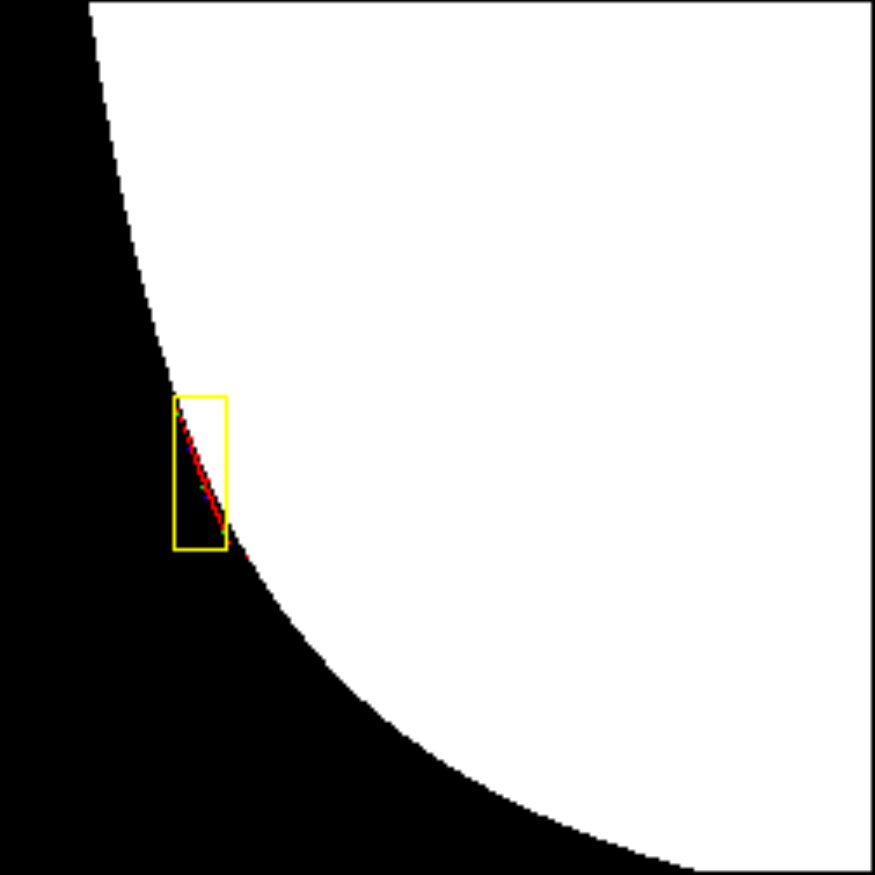}
}
\vspace{1mm}
\hbox{
\includegraphics[width=3.4cm,height=3.4cm]{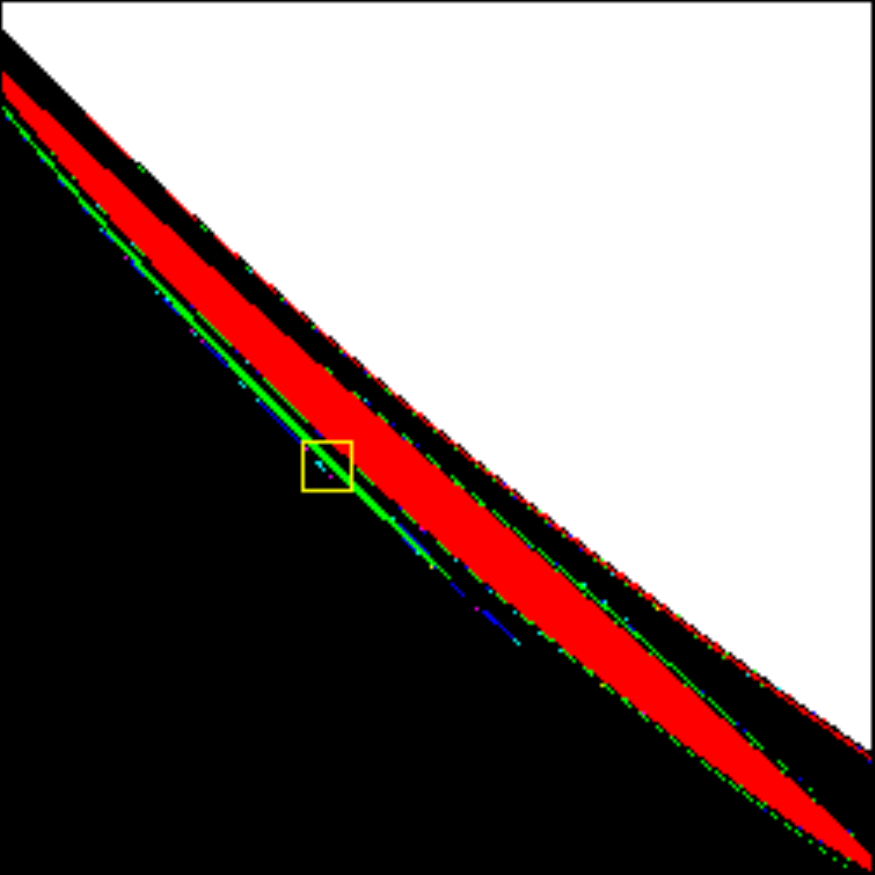}
\includegraphics[width=3.4cm,height=3.4cm]{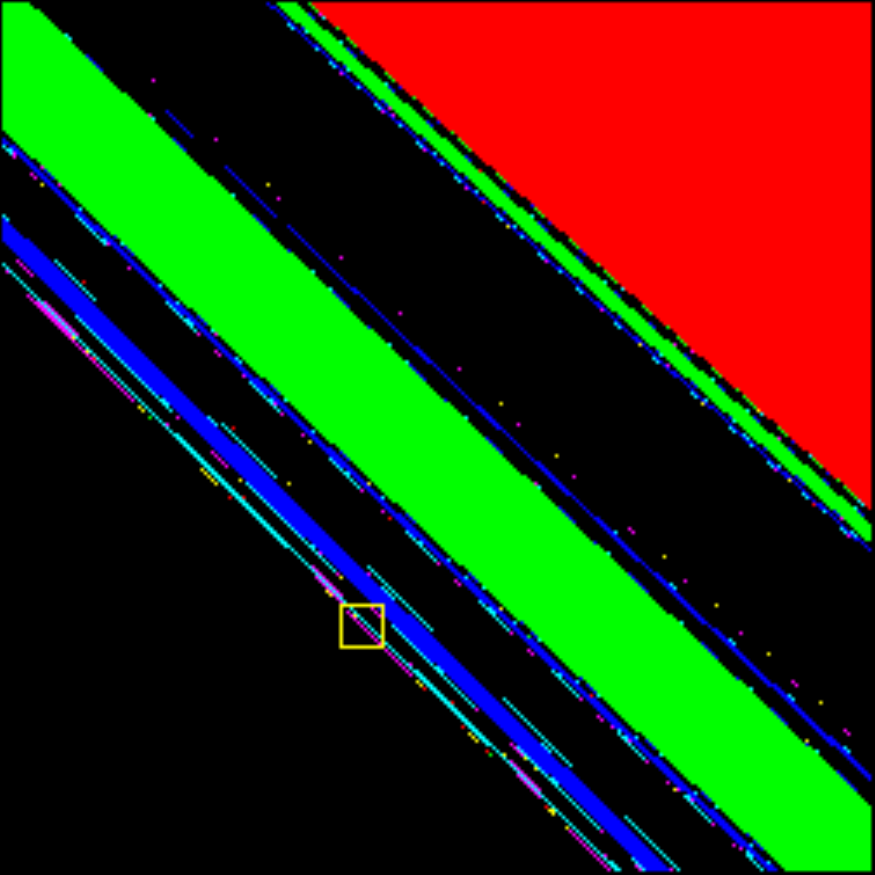}
}
\vspace{1mm}
\hbox{
\includegraphics[width=3.4cm,height=3.4cm]{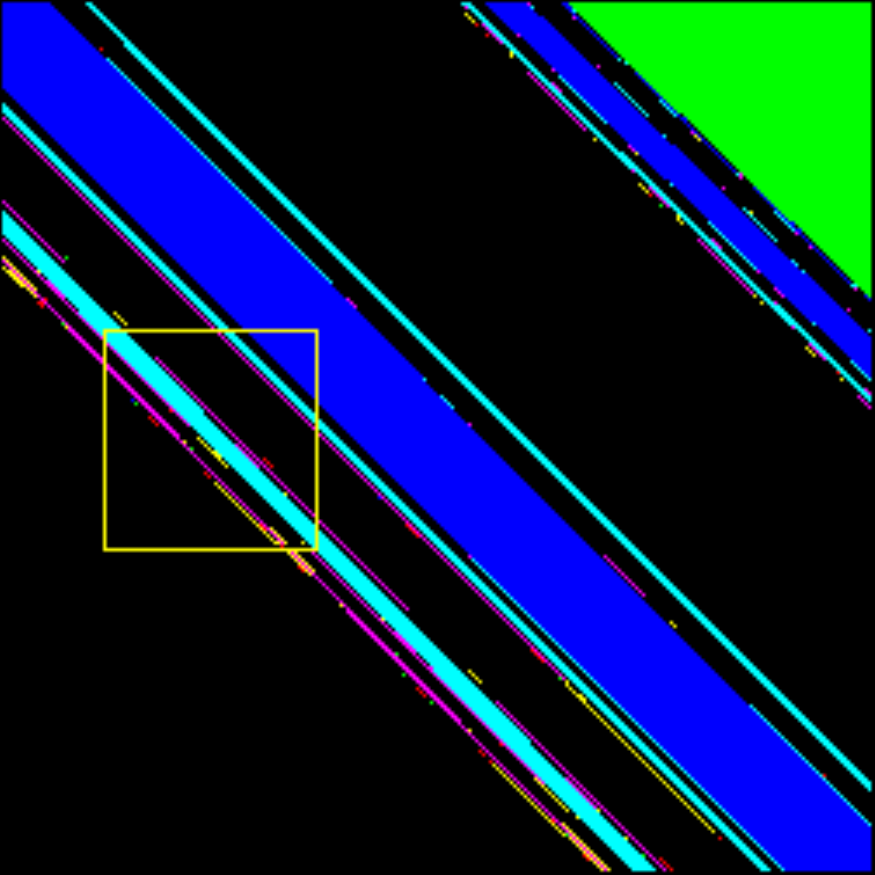}
\includegraphics[width=3.4cm,height=3.4cm]{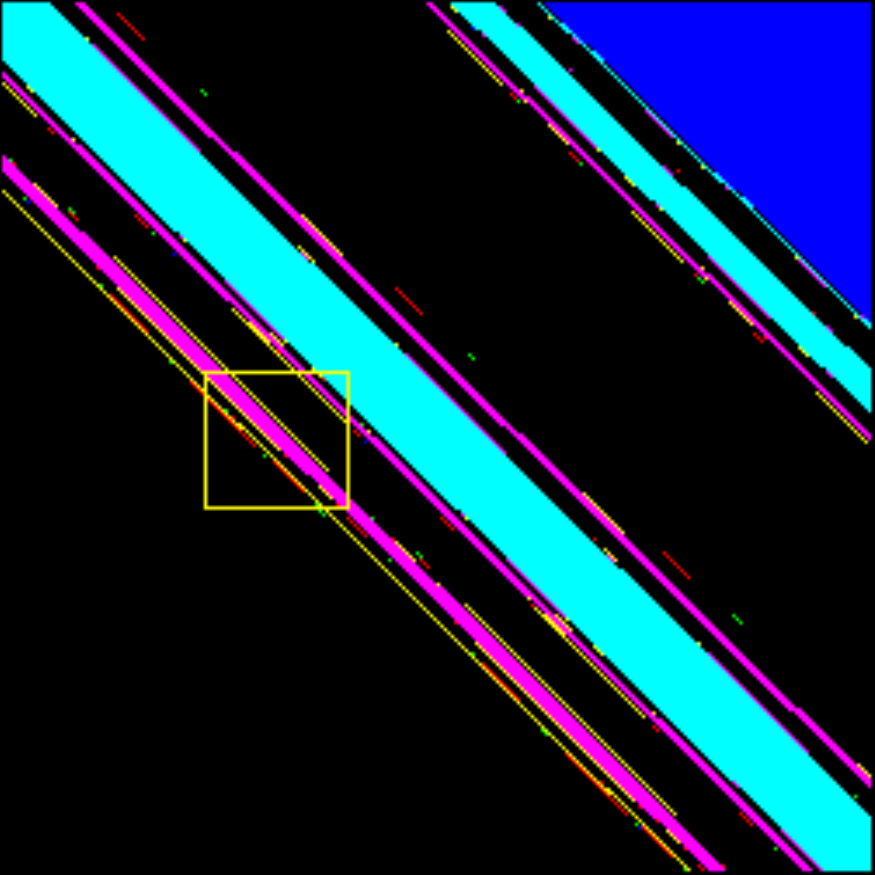}
}
\vspace{1mm}
\hbox{
\includegraphics[width=3.4cm,height=3.4cm]{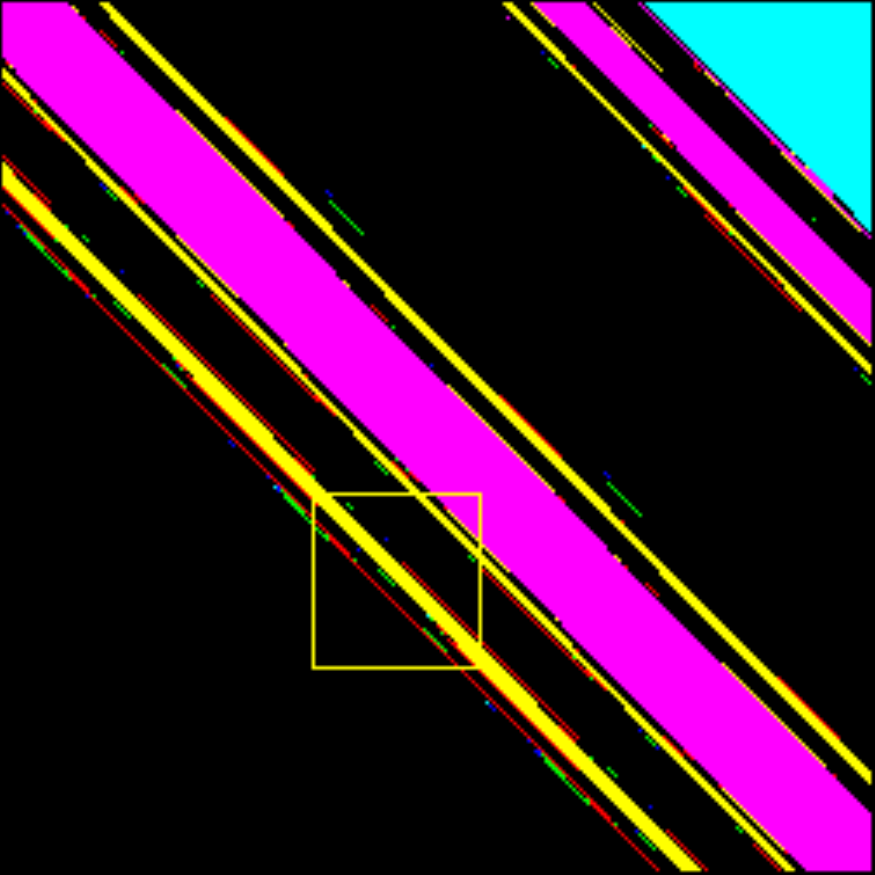}
\includegraphics[width=3.4cm,height=3.4cm]{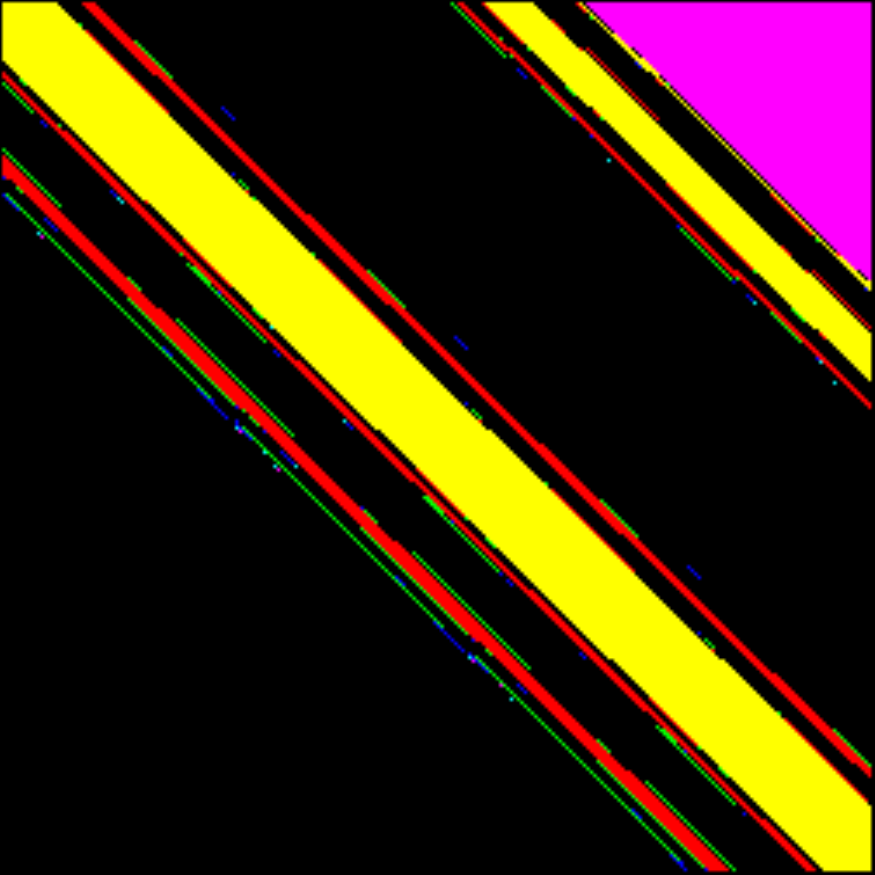}
}
\caption{ (Color online)
Bitmaps of two-dimensional parameter space survey showing the collapse (black region)
or expansion (white or colored regions) of initial data with Gaussian width $\sigma_0$
(horizontal axis) and asymmetry parameter  $\delta$ (vertical axis).
$\mathcal{E}_2$ is used to determine expansion/collapse, and different
shades of grey (colors online) denote PT solutions with different numbers of bounces. 
The top bitmap displays
$1.0 \leq \sigma_0 \leq 7.5$ and 
$0.05 \leq \delta \leq 0.25$,
and the boxes in each bitmap denote the field of view of the next bitmap.
}
\label{fig:FractalBitmap}
\end{figure}
\begin{figure}
\includegraphics[width=8cm,height=8cm]{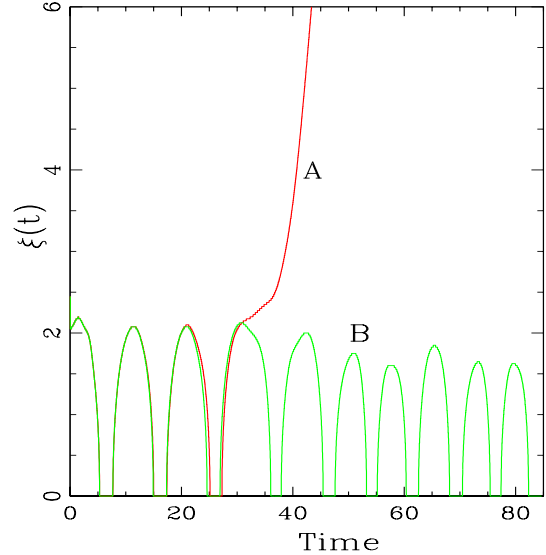} 
\caption{ (Color online)
Plot of $\xi(t)$ for two  solutions subject to criteria $\mathcal{E}_2$.  Solution A is PT but
exits with $N(\xi(t)) = 3$ (i.e., three bounces).
Solution B is NPT and exits due to  the $N(\xi(t)) > N_{\rm osc}$ condition.
}
\label{fig:bounceback}
\end{figure}

An exit criteria that effectively determines which vacuum state the solution will approach
as  $t\rightarrow\infty$ is
\begin{equation}
\mathcal{E}_{2} =  \left\{
\begin{array}{lrrcl}
\textrm{PT} & \rm{when} & \xi(t) & \geq & \xi^+   \\
\\
\textrm{NPT} &  \rm{when} & N(  \xi(t) )& \geq & N_{\rm osc}  \\
                                   &     \rm{or}  &  t & \geq & t_{\rm max} , \\
\end{array}
\right. 
\label{eqn:Condition2}
\end{equation}
where PT and NPT refer to phase transition and no phase transition, respectively;
$\xi^+$ is again the radius beyond which the bubble is assumed to induce a phase transition;
$N_{\rm osc}$ is the number of bounces before assuming the solution enters the oscillon state;
and $t_{\rm max}$ is a time at which the bubble is assumed not to induce a phase  transition
(if $\xi(t)$ has remained less than $\xi^+$).

Figure \ref{fig:exit2} displays three solutions subject to criteria $\mathcal{E}_2$, each exiting due 
to a different condition (denoted A, B, and C in the figure). 
Solution A is an example of a typical runaway PT solution; these
solutions are easily detected with the $\xi(t) \geq \xi^+$
condition and a reasonably chosen $\delta$-dependent  $\xi^+$. 
Solution B is an example of a solution that settles into the oscillon regime, staying localized for
many periods before eventually dispersing (NPT); these solutions are detected with the 
$N(  \xi(t) ) \geq N_{\rm osc}$  condition.
The vast majority of the solutions discussed in this paper are of these two types (A or B).   
The $t\geq t_{\rm max}$ condition is added to the NPT criteria for the cases like solution C, 
where the initial data perturbations are so small that the field immediately disperses
without enough energy to even form an oscillon.  

Figure \ref{fig:FractalBitmap} shows a bitmap displaying the PT/NPT end state as a function 
of $\sigma_0$ and $\delta$ for solutions subject to $\mathcal{E}_2$.
A remarkable difference can be seen when compared to Fig. \ref{fig:simple_scaling}.
The simple boundary between collapsing  and expanding solutions is replaced by a 
much more complicated fractal boundary.
The extra structure  is a result of solutions 
that bounce back after collapse to  form expanding true-vacuum bubbles 
(see Fig. \ref{fig:bounceback}). 
These solutions collapse initially yet still lead to phase transitions.
The solutions shown in Fig. \ref{fig:FractalBitmap} 
are shaded (colored online) differently based on the number of times they bounce before 
expanding to  induce  a phase transition.
Each band of 
$n$ bounces is immediately surrounded by a set of bands of $(n+1)$ bounces that 
approach it from both sides in $\sigma_0$-space.
The structure appears to be fractal given it continues to repeat itself  
in a self-similar fashion upon successive magnification.
This structure closely resembles the structure observed in 
\cite{Matzner_Paper}, except that instead of colliding domain walls (kink/antikink 
soliton collisions) parameterized by boost velocity, it arises from collapsing
bubbles parameterized by initial bubble radius.

The width of $\sigma_0$-space surrounding $\sigma_0^*$ that demonstrates this
fractal bounce behavior is defined to be $\Delta\sigma_0^*$; given this, the 
region of $\sigma_0$-space demonstrating this behavior is given by
\begin{equation}
\Sigma^* = \{ \sigma_0 : (\sigma_0^*-\Delta\sigma_0^*) < \sigma_0 < \sigma_0^* \}.
\end{equation}
In Fig. \ref{fig:FractalBitmap}, it is observed that for $\delta \approx 0.141$,  
$\sigma_0^* \approx 2.52$ and $\Delta\sigma_0^* \approx 0.07$.

\subsection{Threshold of the $n$-bounce band}

After observing the fractal structure of the PT/NPT boundaries in Fig. \ref{fig:FractalBitmap}, 
one might wonder whether there exist scaling laws at the threshold of \textit{each} boundary, 
and if so, whether the scaling law exponents and intermediate attractor solutions 
are universal.
To answer this question, one must be able to determine the boundary 
of a given $n$-bounce band to arbitrary precision.
One might attempt to use  the same bisecting technique used in 
Section \ref{sec:TimeScaling} but with the PT/NPT condition $\mathcal{E}_2$.
Unfortunately, this technique fails due to the fractal nature of the boundary.
For a fixed $\delta$, one cannot use $\mathcal{E}_2$ to find a $\sigma^{PT}_0$ 
(PT end state) and a 
$\sigma^{NPT}_0$ (NPT end state) that bound a single PT/NPT boundary. 
In fact, it appears that there would be an infinite number of such boundaries.
Any bisecting approach using $\mathcal{E}_2$  that began with a $\sigma^{PT}_0$ in the 
desired PT band and a $\sigma^{NPT}_0$ outside the band would almost definitely 
lead to $\sigma^{PT}_0$ hopping to a different band.

However, one can find the boundary of an $n$-bounce band if one exploits the fact that 
an $n$-bounce band is approached by either NPT solutions that form oscillons
or PT solutions with more than $n$ bounces.  
If one starts with a $\sigma_0^{PT}$ that is in the $N_{\rm min}$-bounce band
and a $\sigma_0^{DC}$ that is sufficiently close but not in the band, the 
hybrid condition
%
%
\begin{equation}
\mathcal{E}_3 =  \left\{
\begin{array}{lrrcl}
\textrm{PT} & \rm{when} & \xi(t) & \geq & \xi^+   \\
\\
\textrm{delayed-collapse} &  \rm{when} & N(  \xi(t) )& \geq & N_{\rm min}  \\
                                   &          \rm{and}   &  \xi(t) & \leq & \xi^-  \\
\end{array}
\right.
\label{eqn:Condition3}
\end{equation}
will successfully find the desired edge, denoted $\hat{\sigma}_0^*$,
where $\hat{\sigma}_0^* \in \Sigma^*$
Solutions inside the $N_{\rm min}$-bounce band will induce a phase transition and exit 
due to the $\xi(t) > \xi^+$ condition.
Solutions outside the $N_{\rm min}$-bounce band will bounce more than $N_{\rm min}$ times 
since they either form an oscillon or 
are contained in a PT band with more than $N_{\rm min}$ bounces; when these solutions
bounce more than $N_{\rm min}$  times, they exit as what is here referred to as a 
delayed-collapse solution.  

\begin{figure}
\includegraphics[width=8cm,height=8cm]{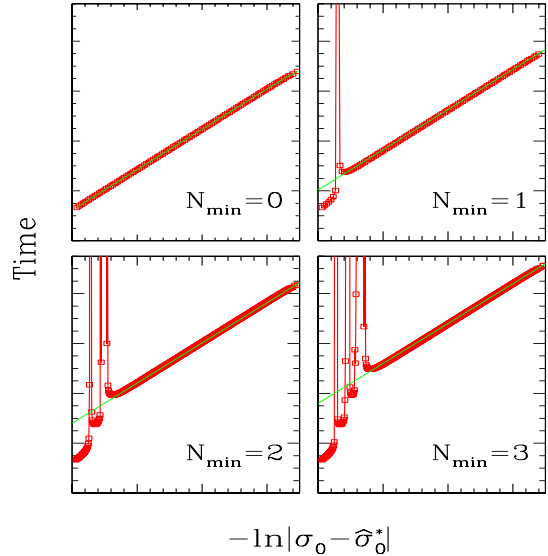}
\caption{ 
Plot of solution lifetime 
as a function of $-\ln| \sigma_0 - \hat{\sigma}^*_0|$ for $\sigma_0$ on the PT side of 
$\hat{\sigma}^*_0$  subject to exit criteria $\mathcal{E}_2$, each 
with different $\hat{\sigma}^*_0$ values obtained by using $\mathcal{E}_3$ with
different $N_{\rm min}$, for  asymmetry parameter $\delta = 0.141$.  
The time-scaling law is observed until the $\sigma_0$ values become large
enough to lie outside the PT band being observed (structure on left side of plots).
}
\label{fig:scaling_all_hi}
\end{figure}
To understand the nature of solutions near the fractal boundary, 
solutions are first evolved with exit criteria $\mathcal{E}_3$ in order to find $\hat{\sigma}^*_0$.  
Then, solutions are obtained with exit criteria $\mathcal{E}_2$ 
using $\sigma_0$ that approach $\hat{\sigma}^*_0$ from either 
side, so that one can observe the
PT or NPT end state of the solutions near $\hat{\sigma}^*_0$.
Fig. \ref{fig:scaling_all_hi} demonstrates the time-scaling law 
\begin{equation}
T = -\gamma \ln | \sigma_0 - \hat{\sigma}^*_0 |.
\label{eqn:TimeScalingHat}
\end{equation}
for $\sigma_0$ approaching $\hat{\sigma}^*_0$ on the PT side.
While similar to the time-scaling seen in Section \ref{sec:TimeScaling},
there now appears to be a scaling law observed on \emph{each} of the 
infinitely many PT/NPT boundaries.  Furthermore, the exponent  $\gamma$ appears to 
be \emph{universal} for a given $\delta$.  
\begin{figure}
\includegraphics[width=8cm,height=8cm]{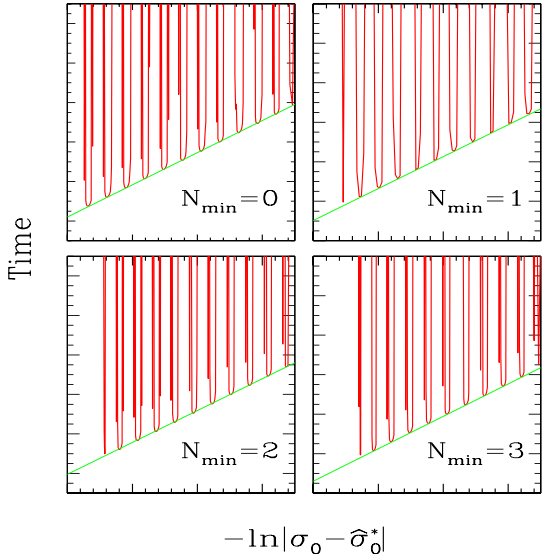}
\caption{ 
Plot of solution lifetime
as a function of $-\ln| \sigma_0 - \sigma^*_0|$ for $\sigma_0$ on the delayed-collapse side of 
$\sigma^*_0$  subject to exit criteria $\mathcal{E}_2$, each
with different $\sigma^*_0$ values obtained by using $\mathcal{E}_3$ with different $N_{\rm min}$,
for  asymmetry parameter $\delta = 0.141$.  
Each $N_{\rm min}$ threshold solution shows roughly equal-width bands of PT solutions 
approaching $\sigma^*_0$ in roughly log-periodic windows of parameter
space.
}
\label{fig:scaling_all_lo}
\end{figure}
Fig. \ref{fig:scaling_all_lo} shows a mix of PT and NPT solutions
for $\sigma_0$ approaching  $\hat{\sigma}^*_0$ from the delayed-collapse side.
The figure seems to imply that each band of PT solutions with 
$n$ bounces is approached by an infinite number of PT bands of 
$(n+1)$ bounces.
For example, in the $N_{\rm min}=1$ plot in Fig. \ref{fig:scaling_all_lo}, each of the 
curves represent PT solutions with $N(\xi(t))\geq 2$ bounces.  
The main dips that can be seen are for PT solutions with $N(\xi(t)) = 2$ bounces;
the regions of fine structure seen around the $N(\xi(t)) = 2$ bands each have $N(\xi(t)) >  2$ bounces,
showing that the structure continues at finer and finer scales.
The gaps between the bands are for $\sigma_0$ that form oscillons and eventually
disperse (with no phase transition).
While the appearance of the bands might only be a truncated fractal structure, the
fact that the bands appear in roughly log-periodic windows that approach $\hat{\sigma}^*_0$ to 
machine precision highly suggests that the bands would continue ad infinitum (Appendix 
\ref{app:MinkowskiDimension}
discusses approximations to the fractal dimension).
The time-scaling law is present yet again (with the same $\delta$-dependent $\gamma$)
but here serves as a lower-limit envelope on the lifetime.  

\begin{figure}
\includegraphics[width=8cm,height=8cm]{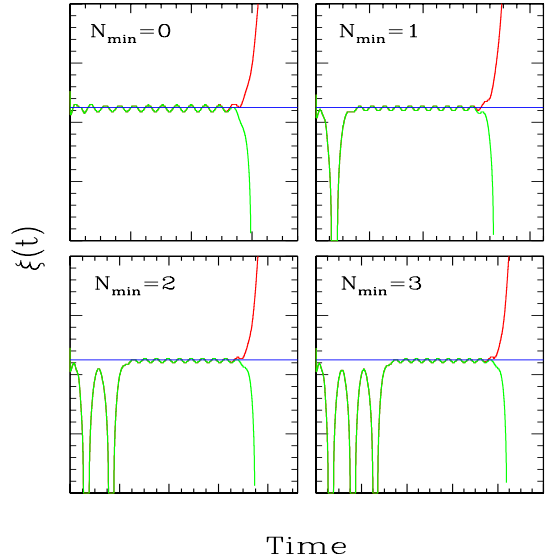}
\caption{ 
Plot of $\xi(t)$ for PT and delayed-collapse solutions for four $N_{\rm min}$ values, 
$\delta=0.141$, and $\sigma_0$ values immediately above and below four distinct critical 
solutions (one for each $N_{\rm min}$).  The horizontal lines in each plot correspond to 
the $\xi$ value for the $\delta=0.141$ static intermediate attractor.  Animations of numerical
evolutions are available online \cite{EPAPS_REFERENCE}. 
}
\label{fig:xi_all}
\end{figure}

Finally, looking at Fig. \ref{fig:xi_all} (animations online, \cite{EPAPS_REFERENCE}) 
one can see that solutions on this
fractal boundary indeed approach the $\delta$-dependent static intermediate attractor.
This strongly suggests that the solution obtained from the static ansatz
(for a given $\delta$) serves as a universal intermediate attractor on the 
threshold of inducing a phase transition.
While the fractal structure itself is not surprising given the previous understanding of kink/antikink
soliton collisions \cite{Matzner_Paper}, 
observing this structure in bubble collapse, the direct measuring of the time scaling law, 
the ability to determine the fractal boundary $\hat{\sigma}^*_0$ to arbitrary precision,
and the realization that the static solution to (\ref{eqn:PhiPrime}) and
(\ref{eqn:phiPrime}) serves as an intermediate attractor on the fractal boundary 
all offer new insight into a classic nonlinear system.


\section{Discussion and Conclusions}

It has been shown that  the boundary between the basins of attraction for the nlKG-ADWP 
system (the true- and false-vacuum states) is fractal in nature.
The fractal structure arises from solutions that collapse, bounce a number of times,
and then expand to induce a phase transition.  
The solutions that induce a phase transition are organized into bands that are 
grouped by the number of bounces they undergo before becoming expanding bubbles.
These bands form a regular and predictable (as opposed to chaotic) fractal structure
in the $\sigma-\delta$ parameter space.   
Bands of solutions with  $n$ bounces are surrounded by bands of solutions with
$(n+1)$ bounces.  When viewed in $\ln|\sigma_0 - \hat{\sigma}^*_0|$-space, the 
roughly equal-width $(n+1)$-bounce bands appear to approach the edges of the $n$-bounce
bands in a log-periodic fashion.
A time-scaling law is observed at each fractal boundary with an exponent that 
depends on the asymmetry parameter $\delta$, and
the solutions that exist on these boundaries are shown to approach the same
static intermediate attractor solution.

While the phenomena discussed here have been studied in the context of isolated scalar 
field collapse, one may wonder what possible impact these phenomena would have
on global phase transitions induced by bubble nucleation.
The Gaussian bubbles (\ref{eqn:FreeSpaceInitialData}) are assumed to represent   
perturbations away from a false vacuum.
For a global phase transition to occur, supercritical perturbations
need to form expanding bubbles that combine, thereby inducing a global transition to the true vacuum.
The impact the fractal behavior discussed in this paper might actually have on bubble nucleation 
would depend not only on how close to threshold these perturbations are for a given 
system, but also on how those perturbations vary throughout space. 
For example, if typical perturbations are near the threshold
of expansion/collapse, $\left< \sigma_0 \right> \approx \sigma_0^*$,
but the variation in $\sigma_0$ in different perturbations throughout space is large,
$\Delta \sigma_0 > \Delta\sigma_0^*$,
the probability that these effects will have an impact on 
previously explored bubble nucleation rates will be low since most $\sigma_0$ are not
contained in $\Sigma^*$ and therefore do not demonstrate the fractal bounce behavior.
However, if $\Delta \sigma_0 \lesssim \Delta\sigma_0^*$, then most $\sigma_0$ are in $\Sigma^*$,
and the phenomena observed in this paper would likely have 
a significant impact on bubble nucleation rates.
In this condition, one might want a way to approximate the probability that a given
perturbation (given by $\sigma_0$) would end up in either the true- or false-vacuum state;
one very natural measure of this probability is related to the fractal dimension itself, 
which is discussed in Appendix A.  
Furthermore, 
one would not need a stochastic description of perturbations
to create a phase transition that appears chaotic or random;  
a smoothly varying distribution of perturbations, where  $\sigma_0\in\Sigma^*$,
would still likely create a system with a rich and complex structure.

\appendix 
\section{Minkowski-Bouligand dimension of log-periodic line segments\label{app:MinkowskiDimension}}

\begin{figure}
\includegraphics[width=8cm,height=8cm]{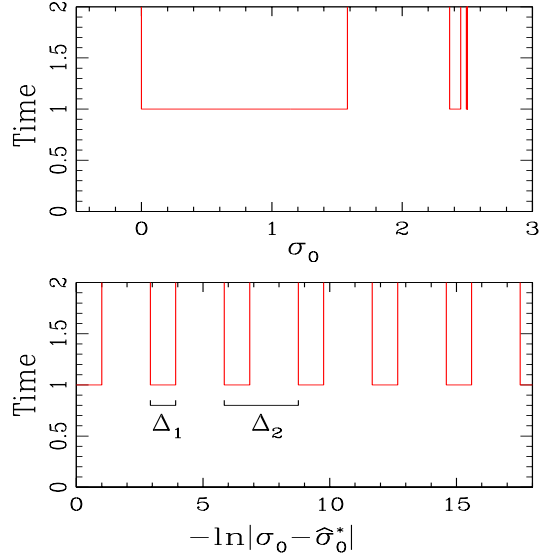}
\caption{ 
Plots demonstrating two observed fractal properties of the PT solutions. 
When viewed in $\ln|\sigma_0 - \hat{\sigma}^*_0|$-space, the PT solutions appear in  
roughly equal-width $(n+1)$--bounce bands that approach the edges of the $n$-bounce
bands in a log-periodic fashion.
The top and bottom plots show  this behavior as a function of $\sigma_0$ 
and
$-\ln|\sigma_0 - \hat{\sigma}^*_0|$, respectively, for $\hat{\sigma}^*_0=2.5$.
}
\label{fig:LogPeriodic}
\end{figure}
%


The structure observed in Fig. \ref{fig:FractalBitmap} seems to be fractal since 
a self-similar banded structure appears at successive scales of magnification. 
For each $n$-bounce band, there appear to be an infinite number of $(n+1)$--bounce 
bands of roughly equal width in $\ln|\sigma_0-\hat{\sigma}^*_0|$-space
that approach the $n$-bounce band
on either side of $\sigma_0$-space in a log-periodic fashion.
By defining a set on $\mathbb{R}^1$ that demonstrates similar behaviors,
this appendix calculates an \emph{approximation} to the fractal 
dimension of the PT/NPT boundary discussed in this paper.

The notion of dimensionality used here is the Minkowski-Bouligand,
or box-counting, dimension,
\begin{equation}
D_M = \lim_{\epsilon\rightarrow 0}
\frac{\log N(\epsilon)}{-\log \epsilon},
\end{equation}
where  $N(\epsilon)$
is the number of boxes of side length $\epsilon$ required to cover the set of PT solutions.
In calculating $D_M$, one can start with a single box covering the region in 
question.  One then successively breaks the box into smaller boxes.  The boxes 
that cover the set contribute to $N(\epsilon)$, and the dimension $D_M$ is obtained in the 
limit $\epsilon \rightarrow 0$.
After fixing $\delta$,  
the $\sigma_0$-$\delta$ boxes reduce to line segments parameterized by
$\sigma_0$.  
Choosing a box that is in the bulk of a PT band yields $D_M = 1$, since every box contains PT solutions
and is in the set, no matter how many times it is divided into smaller boxes.
Choosing a box
that is far away from a PT band and in a clearly NPT region has $D_M = 0$,
since no smaller box ever contains a PT solution allowing it to belong to the set.
The intermediate region is a different story entirely.
\begin{figure}
\includegraphics[width=8cm,height=8cm]{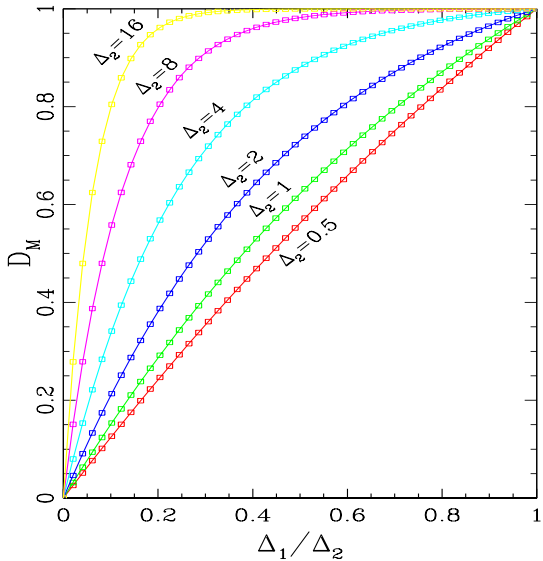}
\caption{ 
Plots of the Minkowski-Bouligand dimension of log-periodic line segments 
for different values of $\Delta_2$ as a function of $\Delta_1/\Delta_2$.  
}
\label{fig:MinkowskiDimension}
\end{figure}

Figure \ref{fig:LogPeriodic} demonstrates the two observed properties of the PT solutions. 
The first property is that the bands of PT solutions appear to be roughly
equal width  in $\ln|\sigma_0-\hat{\sigma}^*_0|$-space, here denoted $\Delta_1$.
The second property is that the bands of PT solutions appear to be an equal
distance, $\Delta_2$, apart in $\ln|\sigma_0-\hat{\sigma}^*_0|$-space. 
The ratio $\Delta_1/\Delta_2$ is essentially a duty cycle quantifying what 
proportion of the $\sigma_0$ space is in the set of PT solutions.
It is this observation that would allow one with limited measurements of $\Delta_1$ and 
$\Delta_2$ to approximate the fractal dimension of the system; 
these numbers could be used to determine the probability that any $\sigma_0\in\Sigma^*$ 
will induce a phase transition.
Figure \ref{fig:MinkowskiDimension} shows the Minkowski-Bouligand 
dimension for different values
of $\Delta_2$ as a function of $\Delta_1/\Delta_2$.  As expected, all solutions have 
$D_M=0$ for $\Delta_1/\Delta_2=0$ since those sets are empty. Likewise, all solutions have 
$D_M=1$ for $\Delta_1/\Delta_2=1$ since those sets are completely solid lines.

The approximation discussed here is only a very rough approximation to the fractal 
dimension observed in the nlKG-ADWP dynamical system since it only considers a single
level of recursion of the self-similar pattern.
In the actual nlKG-ADWP system, in addition to each $n$-bounce window being 
approached by an infinite number of $(n+1)$--bounce windows, each 
$(n+1)$--bounce window also is approached by an infinite number of $(n+2)$--bounce windows, 
and so on.
Therefore, for an observed $\Delta_1$
and $\Delta_2$, the values reported here should be considered a lower bound.

\bibliography{paper}

\begin{thebibliography}{26}
\expandafter\ifx\csname natexlab\endcsname\relax\def\natexlab#1{#1}\fi
\expandafter\ifx\csname bibnamefont\endcsname\relax
  \def\bibnamefont#1{#1}\fi
\expandafter\ifx\csname bibfnamefont\endcsname\relax
  \def\bibfnamefont#1{#1}\fi
\expandafter\ifx\csname citenamefont\endcsname\relax
  \def\citenamefont#1{#1}\fi
\expandafter\ifx\csname url\endcsname\relax
  \def\url#1{\texttt{#1}}\fi
\expandafter\ifx\csname urlprefix\endcsname\relax\def\urlprefix{URL }\fi
\providecommand{\bibinfo}[2]{#2}
\providecommand{\eprint}[2][]{\url{#2}}

\bibitem[{\citenamefont{Bogolubsky and Makhankov}(1976)}]{Bogo_Paper}
\bibinfo{author}{\bibfnamefont{I.~L.} \bibnamefont{Bogolubsky}}
  \bibnamefont{and} \bibinfo{author}{\bibfnamefont{V.~G.}
  \bibnamefont{Makhankov}}, \bibinfo{journal}{JETP Lett.}
  \textbf{\bibinfo{volume}{24}}, \bibinfo{pages}{12} (\bibinfo{year}{1976}).

\bibitem[{\citenamefont{Honda and Choptuik}(2002)}]{Honda_Paper}
\bibinfo{author}{\bibfnamefont{E.~P.} \bibnamefont{Honda}} \bibnamefont{and}
  \bibinfo{author}{\bibfnamefont{M.~W.} \bibnamefont{Choptuik}},
  \bibinfo{journal}{Phys.\ Rev. D} \textbf{\bibinfo{volume}{65}},
  \bibinfo{pages}{084037} (\bibinfo{year}{2002}).

\bibitem[{\citenamefont{Fodor et~al.}(2008)\citenamefont{Fodor, Forgacs,
  Horvath, and Lukacs}}]{GF_PF_ZH_AL_SmallQBs}
\bibinfo{author}{\bibfnamefont{G.}~\bibnamefont{Fodor}},
  \bibinfo{author}{\bibfnamefont{P.}~\bibnamefont{Forgacs}},
  \bibinfo{author}{\bibfnamefont{Z.}~\bibnamefont{Horvath}}, \bibnamefont{and}
  \bibinfo{author}{\bibfnamefont{A.}~\bibnamefont{Lukacs}},
  \bibinfo{journal}{Phys.\ Rev. D} \textbf{\bibinfo{volume}{78}},
  \bibinfo{pages}{025003} (\bibinfo{year}{2008}).

\bibitem[{\citenamefont{Copeland et~al.}(1995)\citenamefont{Copeland, Gleiser,
  and Muller}}]{CGM_Oscillons}
\bibinfo{author}{\bibfnamefont{E.~J.} \bibnamefont{Copeland}},
  \bibinfo{author}{\bibfnamefont{M.}~\bibnamefont{Gleiser}}, \bibnamefont{and}
  \bibinfo{author}{\bibfnamefont{H.~R.} \bibnamefont{Muller}},
  \bibinfo{journal}{Phys.\ Rev. D} \textbf{\bibinfo{volume}{52}},
  \bibinfo{pages}{1920} (\bibinfo{year}{1995}).

\bibitem[{\citenamefont{Gleiser and Haas}(1996)}]{GH_Oscillons_HeatBath}
\bibinfo{author}{\bibfnamefont{M.}~\bibnamefont{Gleiser}} \bibnamefont{and}
  \bibinfo{author}{\bibfnamefont{R.~M.} \bibnamefont{Haas}},
  \bibinfo{journal}{Phys.\ Rev. D} \textbf{\bibinfo{volume}{54}},
  \bibinfo{pages}{1626} (\bibinfo{year}{1996}).

\bibitem[{\citenamefont{Adib et~al.}(2002)\citenamefont{Adib, Gleiser, and
  Almeida}}]{MG_LongLivedAsymmetricBubbles}
\bibinfo{author}{\bibfnamefont{A.~B.} \bibnamefont{Adib}},
  \bibinfo{author}{\bibfnamefont{M.}~\bibnamefont{Gleiser}}, \bibnamefont{and}
  \bibinfo{author}{\bibfnamefont{C.~A.~S.} \bibnamefont{Almeida}},
  \bibinfo{journal}{Phys.\ Rev. D} \textbf{\bibinfo{volume}{66}},
  \bibinfo{pages}{085011} (\bibinfo{year}{2002}).

\bibitem[{\citenamefont{Fodor et~al.}(2009)\citenamefont{Fodor, Forgacs,
  Horvath, and Mezei}}]{GF_PF_ZH_MM_OscillonRadiation}
\bibinfo{author}{\bibfnamefont{G.}~\bibnamefont{Fodor}},
  \bibinfo{author}{\bibfnamefont{P.}~\bibnamefont{Forgacs}},
  \bibinfo{author}{\bibfnamefont{Z.}~\bibnamefont{Horvath}}, \bibnamefont{and}
  \bibinfo{author}{\bibfnamefont{M.}~\bibnamefont{Mezei}},
  \bibinfo{journal}{Phys.\ Rev. D} \textbf{\bibinfo{volume}{79}},
  \bibinfo{pages}{065002} (\bibinfo{year}{2009}).

\bibitem[{\citenamefont{Gleiser and Howell}(2003)}]{MG_RCH_Resonant_2p1}
\bibinfo{author}{\bibfnamefont{M.}~\bibnamefont{Gleiser}} \bibnamefont{and}
  \bibinfo{author}{\bibfnamefont{R.~C.} \bibnamefont{Howell}},
  \bibinfo{journal}{Phys.\ Rev. E} \textbf{\bibinfo{volume}{68}},
  \bibinfo{pages}{065203} (\bibinfo{year}{2003}).

\bibitem[{\citenamefont{Farhi et~al.}(2005)\citenamefont{Farhi, Graham,
  Khemani, Markov, and Rosales}}]{Farhi_SU2_Oscillons}
\bibinfo{author}{\bibfnamefont{E.}~\bibnamefont{Farhi}},
  \bibinfo{author}{\bibfnamefont{N.}~\bibnamefont{Graham}},
  \bibinfo{author}{\bibfnamefont{V.}~\bibnamefont{Khemani}},
  \bibinfo{author}{\bibfnamefont{R.}~\bibnamefont{Markov}}, \bibnamefont{and}
  \bibinfo{author}{\bibfnamefont{R.}~\bibnamefont{Rosales}},
  \bibinfo{journal}{Phys.\ Rev. D} \textbf{\bibinfo{volume}{72}},
  \bibinfo{pages}{101701} (\bibinfo{year}{2005}).

\bibitem[{\citenamefont{Hindmarsh and Salmi}(2006)}]{Hindmarsh_Salmi_2006}
\bibinfo{author}{\bibfnamefont{M.}~\bibnamefont{Hindmarsh}} \bibnamefont{and}
  \bibinfo{author}{\bibfnamefont{P.}~\bibnamefont{Salmi}},
  \bibinfo{journal}{Phys.\ Rev. D} \textbf{\bibinfo{volume}{74}},
  \bibinfo{pages}{105005} (\bibinfo{year}{2006}).

\bibitem[{\citenamefont{Fodor et~al.}(2006)\citenamefont{Fodor, Forgacs,
  Grandclement, and Racz}}]{Fodor_QuasiBreathers2006}
\bibinfo{author}{\bibfnamefont{G.}~\bibnamefont{Fodor}},
  \bibinfo{author}{\bibfnamefont{P.}~\bibnamefont{Forgacs}},
  \bibinfo{author}{\bibfnamefont{P.}~\bibnamefont{Grandclement}},
  \bibnamefont{and} \bibinfo{author}{\bibfnamefont{I.}~\bibnamefont{Racz}},
  \bibinfo{journal}{Phys.\ Rev. D} \textbf{\bibinfo{volume}{74}},
  \bibinfo{pages}{124003} (\bibinfo{year}{2006}).

\bibitem[{\citenamefont{Graham}(2007)}]{NGraham_Electroweak}
\bibinfo{author}{\bibfnamefont{N.}~\bibnamefont{Graham}},
  \bibinfo{journal}{Phys.\ Rev. D} \textbf{\bibinfo{volume}{76}},
  \bibinfo{pages}{085017} (\bibinfo{year}{2007}).

\bibitem[{\citenamefont{Gleiser et~al.}(2008)\citenamefont{Gleiser, Rogers, and
  Thorarinson}}]{MG_BR_JT_BubblingAway}
\bibinfo{author}{\bibfnamefont{M.}~\bibnamefont{Gleiser}},
  \bibinfo{author}{\bibfnamefont{B.}~\bibnamefont{Rogers}}, \bibnamefont{and}
  \bibinfo{author}{\bibfnamefont{J.}~\bibnamefont{Thorarinson}},
  \bibinfo{journal}{Phys.\ Rev. D} \textbf{\bibinfo{volume}{77}},
  \bibinfo{pages}{023513} (\bibinfo{year}{2008}).

\bibitem[{\citenamefont{Gleiser and Thorarinson}(2007)}]{MG_JT_U1}
\bibinfo{author}{\bibfnamefont{M.}~\bibnamefont{Gleiser}} \bibnamefont{and}
  \bibinfo{author}{\bibfnamefont{J.}~\bibnamefont{Thorarinson}},
  \bibinfo{journal}{Phys.\ Rev. D} \textbf{\bibinfo{volume}{76}},
  \bibinfo{pages}{041701} (\bibinfo{year}{2007}).

\bibitem[{\citenamefont{Gleiser and Thorarinson}(2009)}]{MG_JT_HiggsComplexity}
\bibinfo{author}{\bibfnamefont{M.}~\bibnamefont{Gleiser}} \bibnamefont{and}
  \bibinfo{author}{\bibfnamefont{J.}~\bibnamefont{Thorarinson}},
  \bibinfo{journal}{Phys.\ Rev. D} \textbf{\bibinfo{volume}{79}},
  \bibinfo{pages}{025016} (\bibinfo{year}{2009}).

\bibitem[{\citenamefont{Gleiser and Sicilia}(2008)}]{MG_DS_AnalyticOscillons}
\bibinfo{author}{\bibfnamefont{M.}~\bibnamefont{Gleiser}} \bibnamefont{and}
  \bibinfo{author}{\bibfnamefont{D.}~\bibnamefont{Sicilia}},
  \bibinfo{journal}{Phys.\ Rev. Lett.} \textbf{\bibinfo{volume}{101}},
  \bibinfo{pages}{011602} (\bibinfo{year}{2008}).

\bibitem[{\citenamefont{Anninos et~al.}(1991)\citenamefont{Anninos, Oliveira,
  and Matzner}}]{Matzner_Paper}
\bibinfo{author}{\bibfnamefont{P.}~\bibnamefont{Anninos}},
  \bibinfo{author}{\bibfnamefont{S.}~\bibnamefont{Oliveira}}, \bibnamefont{and}
  \bibinfo{author}{\bibfnamefont{R.~A.} \bibnamefont{Matzner}},
  \bibinfo{journal}{Phys.\ Rev. D} \textbf{\bibinfo{volume}{4}},
  \bibinfo{pages}{1147} (\bibinfo{year}{1991}).

\bibitem[{\citenamefont{Honda}(2000)}]{Honda_Diss}
\bibinfo{author}{\bibfnamefont{E.~P.} \bibnamefont{Honda}}, Ph.D. thesis,
  \bibinfo{school}{University of Texas at Austin} (\bibinfo{year}{2000}),
  \eprint{hep-ph/000910}.

\bibitem[{\citenamefont{Choptuik}(1993)}]{Matt_Paper}
\bibinfo{author}{\bibfnamefont{M.~W.} \bibnamefont{Choptuik}},
  \bibinfo{journal}{Phys.\ Rev. Lett.} \textbf{\bibinfo{volume}{70}},
  \bibinfo{pages}{9} (\bibinfo{year}{1993}).

\bibitem[{\citenamefont{Brady et~al.}(1997)\citenamefont{Brady, Chambers, and
  Goncalves}}]{PRB_CMC_SMCVG_Phases}
\bibinfo{author}{\bibfnamefont{P.~R.} \bibnamefont{Brady}},
  \bibinfo{author}{\bibfnamefont{C.~M.} \bibnamefont{Chambers}},
  \bibnamefont{and} \bibinfo{author}{\bibfnamefont{S.~M. C.~V.}
  \bibnamefont{Goncalves}}, \bibinfo{journal}{Phys.\ Rev. D}
  \textbf{\bibinfo{volume}{56}}, \bibinfo{pages}{6057} (\bibinfo{year}{1997}).

\bibitem[{\citenamefont{Choptuik et~al.}(1997)\citenamefont{Choptuik,
  Hirschmann, and Liebling}}]{MWC_EWH_SLL_ApproxBH}
\bibinfo{author}{\bibfnamefont{M.~W.} \bibnamefont{Choptuik}},
  \bibinfo{author}{\bibfnamefont{E.~W.} \bibnamefont{Hirschmann}},
  \bibnamefont{and} \bibinfo{author}{\bibfnamefont{S.~L.}
  \bibnamefont{Liebling}}, \bibinfo{journal}{Phys.\ Rev. D}
  \textbf{\bibinfo{volume}{55}}, \bibinfo{pages}{6014} (\bibinfo{year}{1997}).

\bibitem[{\citenamefont{Choptuik et~al.}(1996)\citenamefont{Choptuik, Chmaj,
  and Bizon}}]{MWC_TC_PB_EYM}
\bibinfo{author}{\bibfnamefont{M.~W.} \bibnamefont{Choptuik}},
  \bibinfo{author}{\bibfnamefont{T.}~\bibnamefont{Chmaj}}, \bibnamefont{and}
  \bibinfo{author}{\bibfnamefont{P.}~\bibnamefont{Bizon}},
  \bibinfo{journal}{Phys.\ Rev. Lett.} \textbf{\bibinfo{volume}{77}},
  \bibinfo{pages}{424} (\bibinfo{year}{1996}).

\bibitem[{\citenamefont{Bartnik and McKinnon}(1988)}]{RB_JM_EYM}
\bibinfo{author}{\bibfnamefont{R.}~\bibnamefont{Bartnik}} \bibnamefont{and}
  \bibinfo{author}{\bibfnamefont{J.}~\bibnamefont{McKinnon}},
  \bibinfo{journal}{Phys.\ Rev. Lett.} \textbf{\bibinfo{volume}{61}},
  \bibinfo{pages}{141} (\bibinfo{year}{1988}).

\bibitem[{\citenamefont{Bizon and Chmaj}(1998)}]{PB_TC_Skyrmions}
\bibinfo{author}{\bibfnamefont{P.}~\bibnamefont{Bizon}} \bibnamefont{and}
  \bibinfo{author}{\bibfnamefont{T.}~\bibnamefont{Chmaj}},
  \bibinfo{journal}{Phys.\ Rev. D} \textbf{\bibinfo{volume}{58}},
  \bibinfo{pages}{041501} (\bibinfo{year}{1998}).

\bibitem[{\citenamefont{Gundlach}(2007)}]{Gundlach_Review}
\bibinfo{author}{\bibfnamefont{C.}~\bibnamefont{Gundlach}}
  (\bibinfo{year}{2007}), \eprint{arXiv:0711.4620v1 [gr-qc]}.

\bibitem[{\citenamefont{Campbell et~al.}(1983)\citenamefont{Campbell,
  Schonfeld, and Wingate}}]{Campbell_Paper}
\bibinfo{author}{\bibfnamefont{D.}~\bibnamefont{Campbell}},
  \bibinfo{author}{\bibfnamefont{J.}~\bibnamefont{Schonfeld}},
  \bibnamefont{and} \bibinfo{author}{\bibfnamefont{C.~A.}
  \bibnamefont{Wingate}}, \bibinfo{journal}{Physica}
  \textbf{\bibinfo{volume}{9D}}, \bibinfo{pages}{1} (\bibinfo{year}{1983}).

\end{thebibliography}

\end{document}